# Microfluidic Thin Film Pressure Balance for the Study of Complex Thin Films

Sébastien Andrieux[1], Pierre Muller[1], Manish Kaushal[1*], Nadia Sofía Macias Vera[1], Robin Bollache[1], Clément Honorez[2], Alain Cagna[3], Wiebke Drenckhan[1]

[1]Université de Strasbourg, CNRS, Institut Charles Sadron UPR22, F-67000, Strasbourg, France
[2]Laboratoire de Physique des Solides, CNRS, Université Paris-Saclay, 91405 Orsay, France
[3]Teclis Scientific, 69380 Civrieux d'Azergues, France

## Abstract

Investigations of free-standing liquid films enjoy an increasing popularity due to their relevance for many fundamental and applied scientific problems. They constitute soap bubbles and foams, serve as membranes for gas transport or as model membranes in biophysics. More generally, they provide a convenient tool for the investigation of numerous fundamental questions related to interface- and confinement-driven effects in soft matter science. Several approaches and devices have been developed in the past to characterise reliably the thinning and stability of such films, which were commonly created from low-viscosity, aqueous solutions/dispersions. With an increasing interest in the investigation of films made from strongly viscoelastic and complex fluids that may also solidify, the development of a new generation of devices is required to manage reliably the constraints imposed by these formulations. We therefore propose here a microfluidic chip design which allows for the reliable creation, control and characterisation of free-standing films of complex fluids. We provide all technical details and we demonstrate the device functioning for a larger range of systems via a selection of illustrative examples, including films of polymer melts and gelling hydrogels.

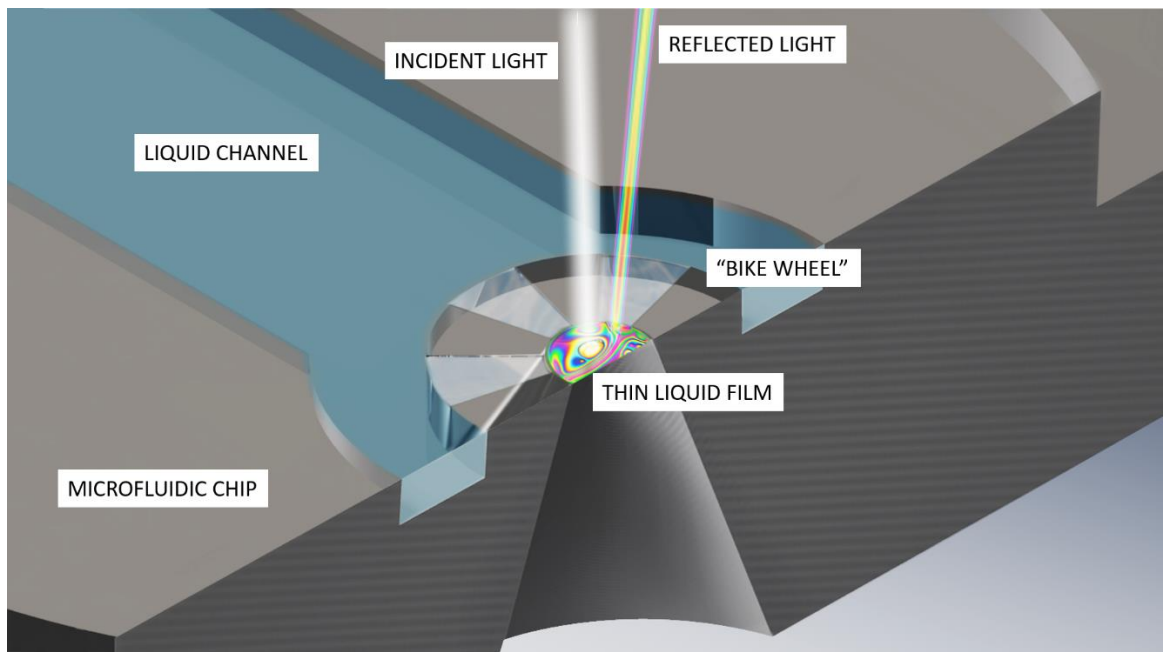

---

* Current address of M.K.: Indian Institute of Technology Kharagpur, Kharagpur-721302, India

## 1. Introduction

Free-standing liquid films are at the core of numerous scientific and applied problems. Since they are thermodynamically metastable, their rupture is commonly delayed by the addition of surface-active agents (low molecular weight surfactants, lipids, amphiphilic polymers, proteins, particles, etc.), which leads to stabilising interactions on the film surface ("interfacial rheology"[1–6]) as well as across the film ("disjoining pressure"[7]). Thanks to this increased stability, extended liquid films of surface areas in the range of mm$^2$-m$^2$ and thicknesses of only 10-1000 nm can be generated. Such free-standing liquid films arise most frequently in the form of soap bubbles and in foams, the latter being essentially three-dimensional arrangements of free-standing films[8]. Due to their large surface-to-volume ratio, liquid films are also increasingly used as membranes for gas separation[9–11]. Apart from practical applications, free-standing liquid films provide excellent model systems for fundamental investigations concerning interface and/or confinement-driven phenomena in soft matter science[12–20].

Due to the above-mentioned importance of free-standing liquid films it is necessary to avail of advanced and easy-to-handle tools to characterise their properties reliably in terms of thinning dynamics and stability. Let us assume that a difference is applied between the pressure $p_g$ of the gas surrounding the film and the pressure $p_l$ of the liquid in the film, as sketched in Figure 1a. This pressure difference drives film thinning ("drainage") and is commonly called the capillary pressure: $p_c = p_g - p_l$. In many cases, it arises naturally due to gravity. The main four stages of the evolution of such a film are then:

(1) The drainage of an initially thick film (typically with a film thickness $h$ > 100 nm) is fully governed by hydrodynamics. The viscoelastic monolayer of stabilising agents at the film surface imposes boundary conditions on the flow[1,17,21].
(2) The film becomes thin enough ($h$ < 100 nm) so that the two monolayers of stabilising agents start to interact via DLVO forces (electrostatics, van der Waals, steric…) across the film[15], creating a disjoining pressure $\Pi$ (Figure 1b), which can either promote ($\Pi$ < 0) or hinder ($\Pi$ > 0) drainage. The confinement may also create structural forces in the bulk of the film, leading, for example, to stratified drainage[22–26].
(3) When $\Pi = p_c$, an equilibrium is reached, and film drainage stops. Different capillary pressures $p_c$ lead to different film thicknesses, allowing therefore to establish disjoining pressure isotherms $\Pi(h)$ to quantify the DLVO interactions in the film[15].
(4) After a certain time and/or beyond a certain disjoining pressure $\Pi$, the film ruptures[27,28].

The initial drainage stage is often investigated via gravity-driven drainage[1,17,21] of vertical films [1,21,29] or on individual bubbles[30]. However, this renders an explicit control of the capillary pressure difficult and other effects become important (marginal regeneration, stability of the film under its own weight, etc.). Therefore, a whole class of devices - so-called "Thin Film Pressure Balances" (TFPB)[15] - has been developed to investigate horizontal films under an explicitly applied capillary pressure. The main functioning of all TFPBs is identical: a controlled capillary pressure is applied to drain the film down to its equilibrium state (and potentially rupture). Optical interference techniques with white or monochromatic light are used to determine the film thickness (profile or point measurement) during drainage as well as at equilibrium. However, the devices differ in (i) how the film is held, and (ii) how the capillary pressure is applied.

Concerning (i), the historical Scheludko cell (Figure 1c.I)[31–33] contains a millimetric film at ambient gas pressure whose liquid pressure is controlled via a liquid channel. The maximum capillary pressure that can be reached is of the order of the entry pressure of the gas into the liquid channel, i.e. inversely proportional to the channel width. This leads to characteristic capillary pressures of the order of $10^2$ Pa, which is often too low to investigate relevant phenomena. Moreover, the asymmetric feeding of the film can lead to non-negligible dynamic effects. Therefore, Mysels proposed in 1964 the use of a porous glass ring[34,35] with very fine pores (hence a large entry pressure) to hold the film. This approach was later improved by Exerowa and Scheludko[36] (Figure 1c.III) via the use of a glass disk, whose conical hole holds the film solidly in place without the formation of a large meniscus. This Exerowa-Scheludko cell has since then become one of the standard techniques, allowing to work with very homogeneous film feeding at capillary pressures up to many thousands of Pa[37,38]. However, the use of porous discs has also disadvantages, such as adsorption phenomena of the surfactants due to the high surface to volume ratio of the porous disc, clogging, and difficult cleaning. This is why Radke[39] proposed in 2001 a compromise between both film holders in form of a "bike wheel" (Figure 1c.II) feeding the film with 24 evenly distributed lithography-produced channels of less than 100 µm thickness. The same approach was successfully applied in 2016 by Beltramo et al.[40]

Concerning (ii), many different approaches exist to apply a well-controlled capillary pressure in the chosen film holder. The simplest approach is to work at ambient gas pressure and to control the liquid pressure via a hydrostatic pressure feed. More sophisticated approaches control the liquid volume with a syringe pump and measure the resulting pressure, or use directly a pressure controller[25,41–43]. Other techniques apply a constant hydrostatic pressure to the liquid and control explicitly the gas pressure[22,29,44].

The described approaches have been proven very useful in advancing our understanding of free-standing films generated from solutions/dispersions of reasonably low viscosities[12–20]. However, TFPBs are increasingly used for the investigation of strongly viscoelastic, complex fluids, including fluids undergoing solidification/gelation. This puts new constraints on the technique, requiring a highly flexible and rapid management of high pressures and the ease to clean the device after use. Other constraints are added naturally, such as the need to replace the gas phase on either side of the film or to replace the gas by a liquid phase to investigate liquid/liquid films.

In order to find a solution for such a versatile device, we developed a microfluidic TFPB ("µTFPB"), whose core is a microfluidic Radke-style bike wheel with short feeding channels entirely integrated into a three-dimensional millifluidic chip. The geometrical flexibility in the design of this part allows to adapt the microfluidic channel dimensions (width and length) to find the appropriate compromise between two conflicting parameters: maximum entry pressure and minimum flow resistance. One of the main advantages of this chip is that it is composed of several layers, containing different ensembles of channels and windows. These layers can be separated and re-assembled easily for cleaning and repeated measurements. The chip geometry and layer assembly are described in detail in Section 2.1. Both, gas and liquid pressure are explicitly set using computer-controlled pressure controllers, as detailed in Section 2.2. In contrast to existing devices, this provides a highly responsive and flexible control of the applied capillary pressure with a variable pressure offset. Section 2.3 presents the optical system used to visualise and quantify the film thickness, which allows for the study of film drainage.

The goal of this article is to provide to the community a detailed description of the developed µTFPB and show the range of fluids accessible for study using a selection of illustrative examples, including a simple surfactant solution (Section 3.3.1) and a polymer melt (Section 3.3.2). We also show for the first time the use of a TFPB for solidifying hydrogel films (Section 3.3.3).

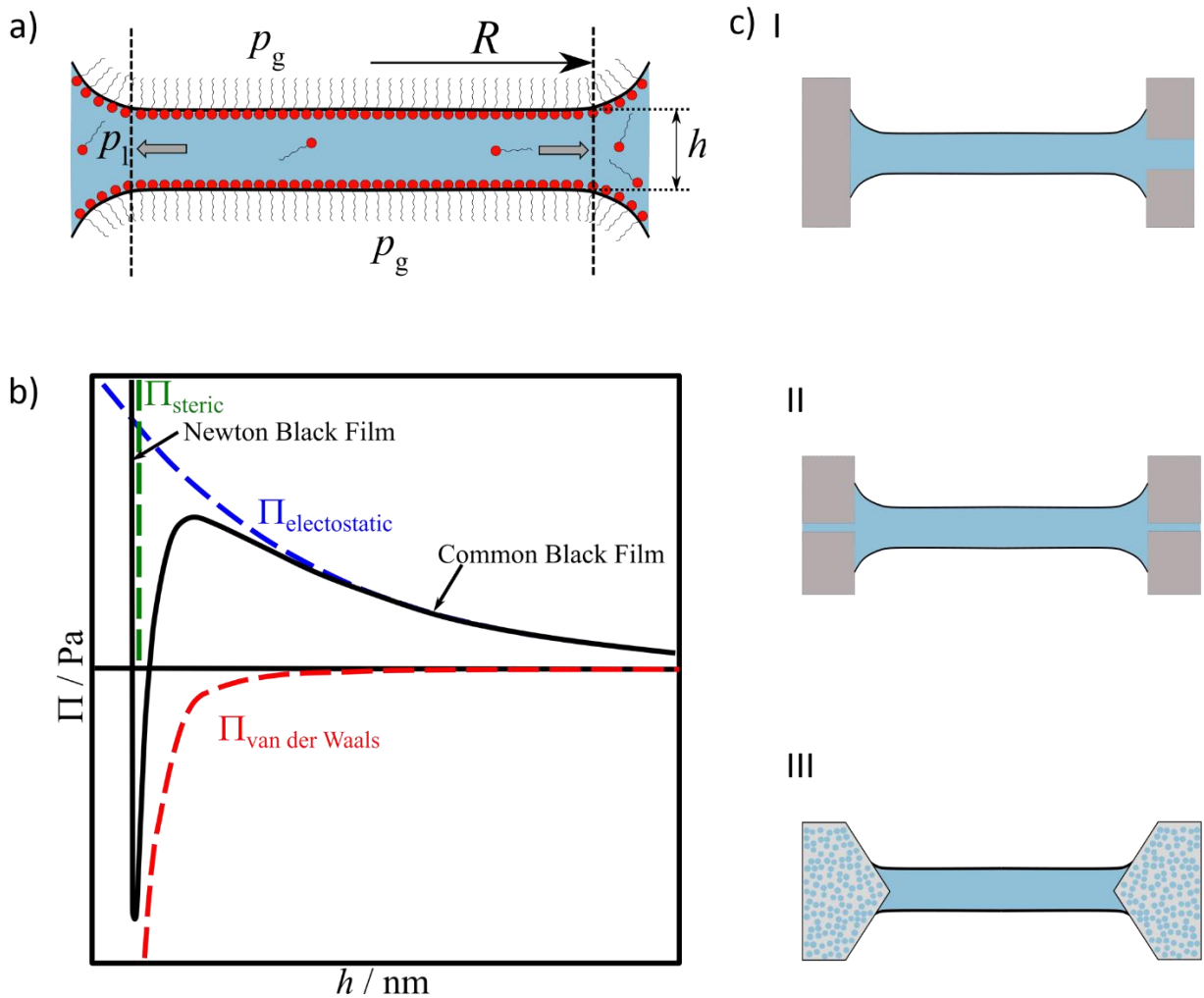

**Figure 1:** (a) Sketch of a free-standing film of thickness $h$ stabilised by surface active agents (not drawn to scale) at a capillary pressure $p_c = p_g - p_l$. The grey arrows show the liquid flow out of the film. (b) The different contributions of the disjoining pressure $\Pi$ (without bulk contributions), redrawn from[15]. (c) The three main approaches of holding and feeding the film: (I) Scheludko cell, (II) Radke's bike wheel, and (III) Exerowa-Scheludko cell (porous disc).

## 2. Design and functioning of the microfluidic Thin Film Pressure Balance

The main innovation of the microfluidic thin film pressure balance (µTFPB) is the microfluidic chip. It is integrated into the overall set-up as sketched in Figure 2a and b. The set-up can be divided into three main parts: the microfluidic chip (Section 2.1), the fluid control (Section 2.2) and the optical system (Section 2.3).

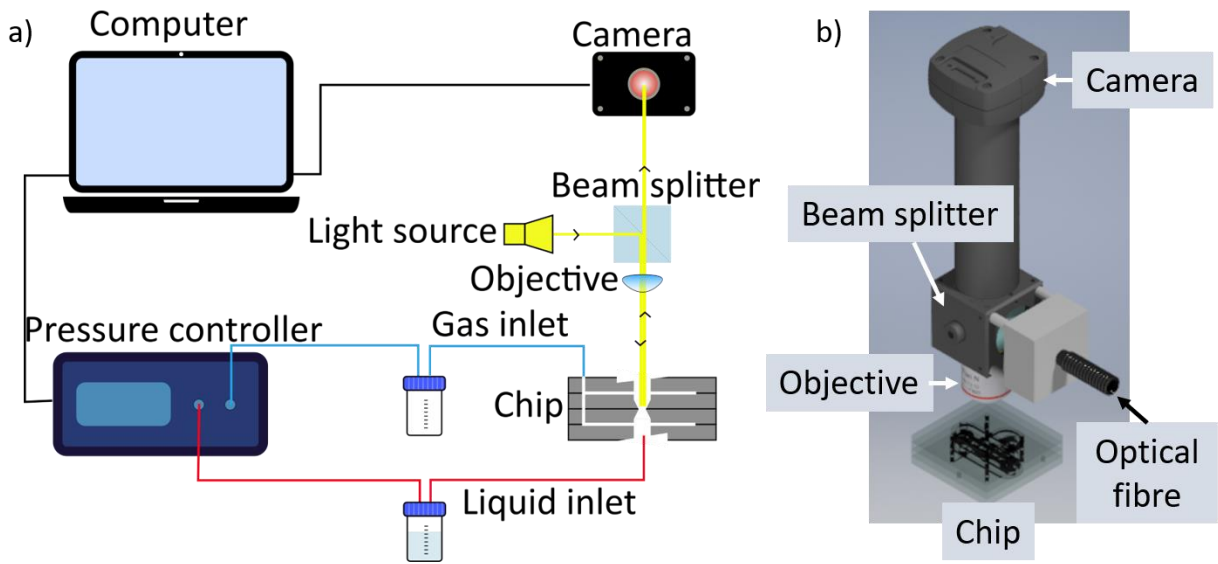

**Figure 2:** a) General setup of the microfluidic thin film pressure balance: a multi-channel pressure controller applies a well-defined capillary pressure to a free-standing film maintained by a bike-wheel type cone in the centre of the microfluidic chip. The film is visualised (and its thickness analysed) using interference of white light reflected from the film and detected by a CCD camera. b) Three-dimensional view of the chip and the optical system.

### 2.1. Microfluidic chip

The microfluidic chip constructs a three-dimensional channel network to create, control and analyse a millimetric, horizontal liquid film created in its centre. The chip is composed of four plates (5x5 cm, numbered from A to D, in Figure 3a), each one being independently milled in COC (Cyclic Olefin Copolymer, commercial name: TOPAS 8007X4, 4 mm-tick wafers, from Microfluidic Chip Shop) with a micromilling machine (SLS Micro Mill from Minitech). **Erreur ! Source du renvoi introuvable.** in the supplementary materials summarises the tools used to mill the different plates constituting the chip, and the parameters used for milling. The channel geometry is designed using Inventor® (computer-aided design software from Autodesk). The software Esprit® (computer-aided manufacturing software from DP Technology) is used to translate the design into the final data format used for the milling process.

Gas- and liquid-tight sealing between the different plates of the chip is ensured by commercial rubber joints (1.6 mm thick, black rings in Figure 3c,d) positioned in specifically designed channels (width: 2.00 mm, depth: 1.28 mm). The plates are held together tightly by four bolts and nuts. The two outer parts of the chip (A and D) are milled to obtain an inclined surface with a 5° angle to avoid any parasite reflection from the chip that could disturb image analysis. 3 mm wide holes are also drilled into these plates above and below the place where the film forms to avoid any loss of intensity and leave a clear path for the light. The inclined, drilled surfaces, one of which can be seen on plate D in Figure 4a, are sealed by gluing a microscope square cover slip (see Figure 3c,d). Alignment issues of the plates may arise, yielding disturbing shades and reflections upon observation of the film. In such a case, it may be useful to add guiding bolts into holes drilled at the exact bolt diameter using the micromilling machine.

The film is formed between the parts B and C, at the intersection of two cones (see Figure 3b and f), so that the film has a diameter of 1 mm. The position of the bike wheel is pointed at by the green arrow on plate B in Figure 3a. The liquid arrives at the hole between the two cones from the side and enters the chip from the bottom of plate A, from the bottom. The gas is injected in the inlet on plate D and circulates on both sides of the film via connecting channels between the different parts of the chip (see the blue arrows in Figure 3b), ensuring an equal pressure on both sides of the film.

The liquid film forms within a bike wheel-type cell (see Figure 3f), whose geometry we modified with respect to the one proposed by Radke[39] to optimise for the study of viscous liquids at high capillary pressures. The film is fed homogeneously via six channels with an entry width of 259 µm and a height of 20 µm. The small channel height ensures an entry pressure of the order of 10000 Pa. Moreover, the channels are only 1 mm long to reduce the pressure drop required to push the liquid. The channels are fed homogeneously by a much larger and deeper ring (1 mm wide and 500 µm deep). All channels and tubing leading to the bike wheel are at least 1 mm wide. Hence the main pressure drops occurring when pushing the liquid through the chip arises in the channels of the bike wheel. This part of the chip can be easily adjusted for different liquids in order to find the appropriate compromise between maximising the entry pressure and minimising the hydrodynamic resistance of the flow. Another important advantage of this plate configuration in contrast to previous designs is that it can be opened easily in a manner that all channels are fully accessible for cleaning. The chip can be easily re-assembled for repeated experiments as long as liquids are used for the experiments and the cleaning liquid do not damage the COC.

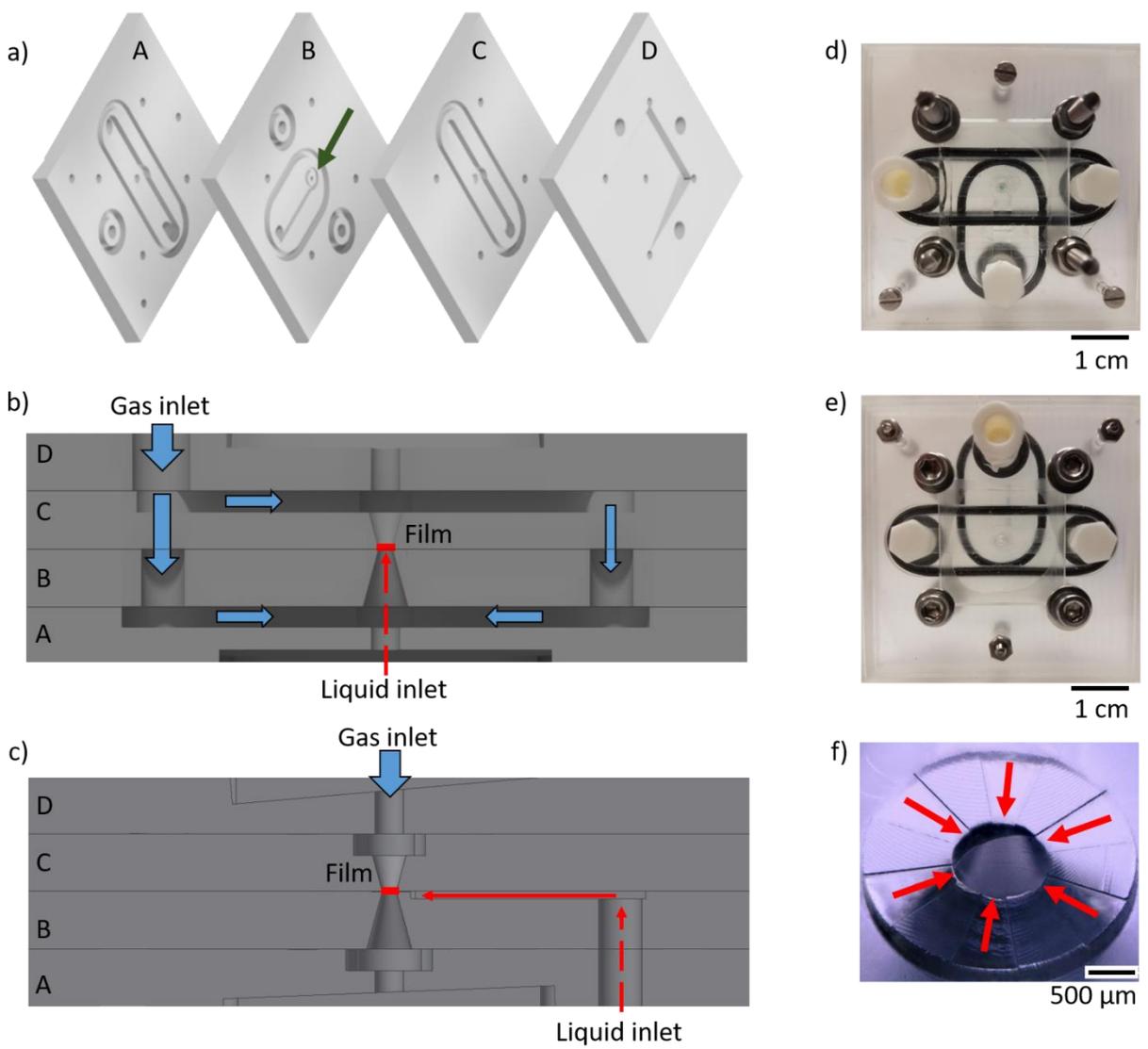

**Figure 3:** Schematic representation of the microfluidic chip a) divided in its four components, with the green arrow showing the position of the bike wheel on part B and b,c) its assembly from two different orientations, showing where the gas (blue arrows) and the liquid (red arrows) flow. b) and c) are rotated by 90°. Picture of the assembled microfluidic chip: d) top view, showing the gas inlet; e) bottom view, showing the liquid inlet[1]. f) Photograph of the bike wheel, with the red arrows showing the motion of liquid into the wheel to form the film.

2.2. Fluid control

Both the liquid and the gas pressure are controlled via a multi-channel pressure pump OB1 Mk3 from Elveflow with a pressure range of 0 to 2000 Pa (Figure 2a). In contrast to previous device designs, this provides a highly responsive application of the capillary pressure with a tuneable pressure offset. The latter controls the pressure difference with the atmospheric pressure, which can be an important parameter for pressure-sensitive applications (gas mixtures, etc.). If desired, a pressure controller with a higher maximal pressure can be used (at the expense of precision in the low-pressure range). We tested that the chip remains perfectly sealed up to $10^5$ Pa by observing an absence of bubble formation when the chip was held under water. Beyond this range, the capillary pressure overcomes the entry pressure, leading to the entry of gas into the liquid channel. A full characterisation of films needs to combine pressure controllers of different ranges to access a wide pressure range with the required precision, especially in the low-pressure range.

---

[1] Some extra inlets which are sealed are visible in Figure 3d,e). These are remnants of former chip designs, but we left the holes for potential alternative inlets: the chip being close to the objective, some connectors may need to be moved to avoid overload issues.

The liquid is stored in a bottle with a GL45 cap linked on one side to the pressure controller and on the other side to the microfluidic chip. The liquid is pushed into the chip by applying a pressure on the bottle. A similar bottle is also inserted between the gas outlet of the pressure controller and the gas inlet of the microfluidic chip. The use of this bottle is three-fold. (1) It protects the pressure controller from being flooded by the liquid in case of a sudden pressure increase at the liquid inlet. (2) It allows for saturating the atmosphere within the chip with the solution by depositing the liquid at the bottom of the bottle[45]. The chip is flooded with humid gas before plugging the liquid channel connector in. (3) It homogenises pressure fluctuations.

Both the liquid and gas pressures are set using the Elveflow Smart Interface software (from Elveflow), providing straightforward control over the capillary pressure and hence the film drainage and equilibrium thickness. However, one must pay attention to the liquid pressure $p_l$, which is not equal to that set by the pressure controller. One needs to add the hydrostatic pressure difference $\Delta p_H = \Delta\rho g H$, where $H$ is the height difference between the film and the surface of the solution in the bottle. $\Delta\rho$ is the density difference between the solution and the gas, and $g$ is the acceleration of gravity. To avoid this correction, it is convenient to place the bottle on a lab jack to align the level of liquid in the bottle with the centre of the microfluidic chip, i.e. $H = 0$.

For solutions with low viscosity, one may simply use the hydrostatic pressure to fill the chip and form the film without having to apply an additional pressure with the pressure controller. Changing the height of the liquid bottle suffices to form a soap film similar to the classical TFPB. Yet, for highly viscous solutions, the hydrostatic pressure and the pressure applied by the pressure controller may not be enough to push the liquid into the chip. One may also use a syringe pump instead to push the liquid and form the film. This requires, however, the integration of an additional pressure sensor to monitor the liquid pressure $p_l$.

### 2.3. Optical System and film thickness measurement

The analysis of the film thickness is done via interferometry (Figure 2b). The procedure described here is fairly standard for a thin film pressure balance and has been validated over the last 50 years of research[46]. In our device, the film is homogeneously exposed to white light from a halogen lamp (15 V, 150 W). The lamp illuminates the chip via an optical fibre (shown in Figure 2b) directed towards a beam splitter which redirects the light towards the chip, as schematised by the yellow arrow in Figure 2a. A 4x objective focuses the light to the area where the film forms. The reflected light passes again through the objective and the beam splitter before being collected by a 12 bit CCD camera (UEye UI-3580LE-C-HQ from iDS Imaging Development Systems GmbH) with a CMOS colour sensor and a resolution of 4.2 Mpx. The ensemble is constructed using Thorlabs elements combined with 3D-printed pieces to hold the light source and the microfluidic chip. Alignment of the film with the optical path is done using a combination of a mechanical translation and rotation stage. A monochromatic filter can be used in front of the camera to obtain monochromatic images for thickness measurements. However, thanks to the quality of modern cameras it is possible to profit from the RGB splitting of unfiltered images instead without loss in precision[24–26]. Moreover, it has the advantage of availing of fully coloured images for general visual inspection of the film in parallel to the possibility of quantitative data analysis.

For this purpose, the images are treated with the open-source software ImageJ. First, we split the channels using the RGB splitting tool. This procedure provides three grey-scale images. All three colours can be used for treatment and even be combined for better precision[24–26]. We treat here only the red channel, since it provides the best precision due to its longer wavelength. This choice needs to be reconsidered, however, if strongly absorbing liquids are used. We use λ = 600 nm for the wavelength since it corresponds to the maximal sensitivity of the camera for the red channel.

Interference in liquid films is periodic with $\lambda/2n_l$ with the first constructive maximum in reflected intensity $I_r^{max}$ arising at $h = \lambda/4n_l$ (113 nm for water), where $n_l$ is the refractive index of the liquid. Here we are interested in quantitative measurements only for film thicknesses below this value.

We select a circular region of interest (ROI) on the film with an area of 1024 pixels, in which we measure the average grey value $I_r$. This operation is repeated on all the images recorded for the film using an ImageJ macro. We calculate the film thickness $h$ at each time $t$ for the same ROI using the Scheludko equation[46]

$$h = \frac{\lambda}{2\pi n_l} \arcsin \sqrt{\frac{\Delta}{1 + \frac{4R}{(1-R)^2}(1-\Delta)}},$$

with λ the wavelength, $R$ the reflexion coefficient given by $R = [(n_l - n_g)/(n_l + n_g)]^2$, where $n_g$ is the refractive index of the gas. Δ is the relative intensity which is defined using the maximal reflected intensity $I_r^{max}$ and the minimal reflected intensity $I_r^{min}$

$$\Delta = \frac{I_r - I_r^{min}}{I_r^{max} - I_r^{min}}.$$

The maximal reflected intensity $I_r^{max}$ is obtained during the early stages of film thinning at $h = \lambda/4n_l$, when the film passes through fully constructive interference, while the minimal reflected intensity $I_r^{min}$ is measured when the film is broken. $I_r^{min}$ is a collection of residual intensity resulting from the environment and different reflections within the set-up etc. Since this is a constant contribution, it can be subtracted on all images.

For reliable image analysis, the camera settings (in particular the exposure time) is chosen such that the obtained grey values cover a maximally wide range for best precision. Moreover, the gray values should be in a range between 50 and 200 where the captor sensitivity is highly linear.

## 3. Examples
In this section we demonstrate examples of thin films studied with the µTFPB. The first example (Section 3.3.1) is a film made of an SDS (sodium dodecylsulfate) solution at a concentration of 0.24 mM (i.e. 3 times the critical micellar concentration, without any added salt). The viscosity of the SDS solution is that of water ($10^{-3}$ Pa s). The second example (Section 3.2) is a film made from a polymer melt (DBP-732 from Gelest) consisting of comb polymers with a polydimethylsiloxane backbone with side chains of poly(ethylene glycol)/poly(propylene glycol) copolymers). DBP-732 has a significantly higher viscosity (2.7 Pa s[47]) compared to the SDS solution, which makes the study of films very difficult with a classical TFPB, as discussed in Section 1. Finally, in Section 3.3 we show that the µTFPB allows us to monitor the drainage and stability of a film during solidification, using the example of a physically gelling alginate hydrogel. The examples chosen are illustrative of the range of complex fluids which can be studied with this µTFPB and do not aim to shed light on the hydrodynamics and physical chemistry involved in the drainage of these systems.

### 3.1. Low viscosity foam films (0.24 mM SDS solution)

Figure 4.I.a-d shows the drainage over time of a 0.24 mM SDS solution (viscosity $10^{-3}$ Pa s) subjected to a capillary pressure of 100 Pa monitored with the µTFPB. The different colours and grey levels correspond to different film thicknesses. Figure 4.I.a-d shows the evolution of the film thickness $h$ over time in the region depicted by the yellow circle pointed by the white arrows in the corresponding images using the Scheludko equation (Section 2.3). We used the refractive index of water, i.e. $n_l$ = 1.33, to obtain the film thickness. One notices a rapid film thinning down to an equilibrium thickness of 24 ± 2 nm. One also observes the well-known stratification phenomenon, with the strata being marked by the grey dotted lines in Figure 4.I.e, typical of surfactant films above the cmc. Such stratification originates from the three-dimensional packing of the electrically charged micelles within the film. The micelles leave the film layer by layer, yielding regions of discrete thicknesses depending on the number of micellar layers in the film[15,16,48]. The thickness of each step corresponds to the characteristic distance between micelles fixed by its electrostatic double layer, which here corresponds to 10 ± 3 nm. The equilibrium thickness of ca. 24 nm is reached after 180 s of drainage, which is in agreement with the literature[16,49]. This experiment thus shows that the µTFPB is indeed as efficient as the previous thin film pressure balances for the study of films of low viscosity.

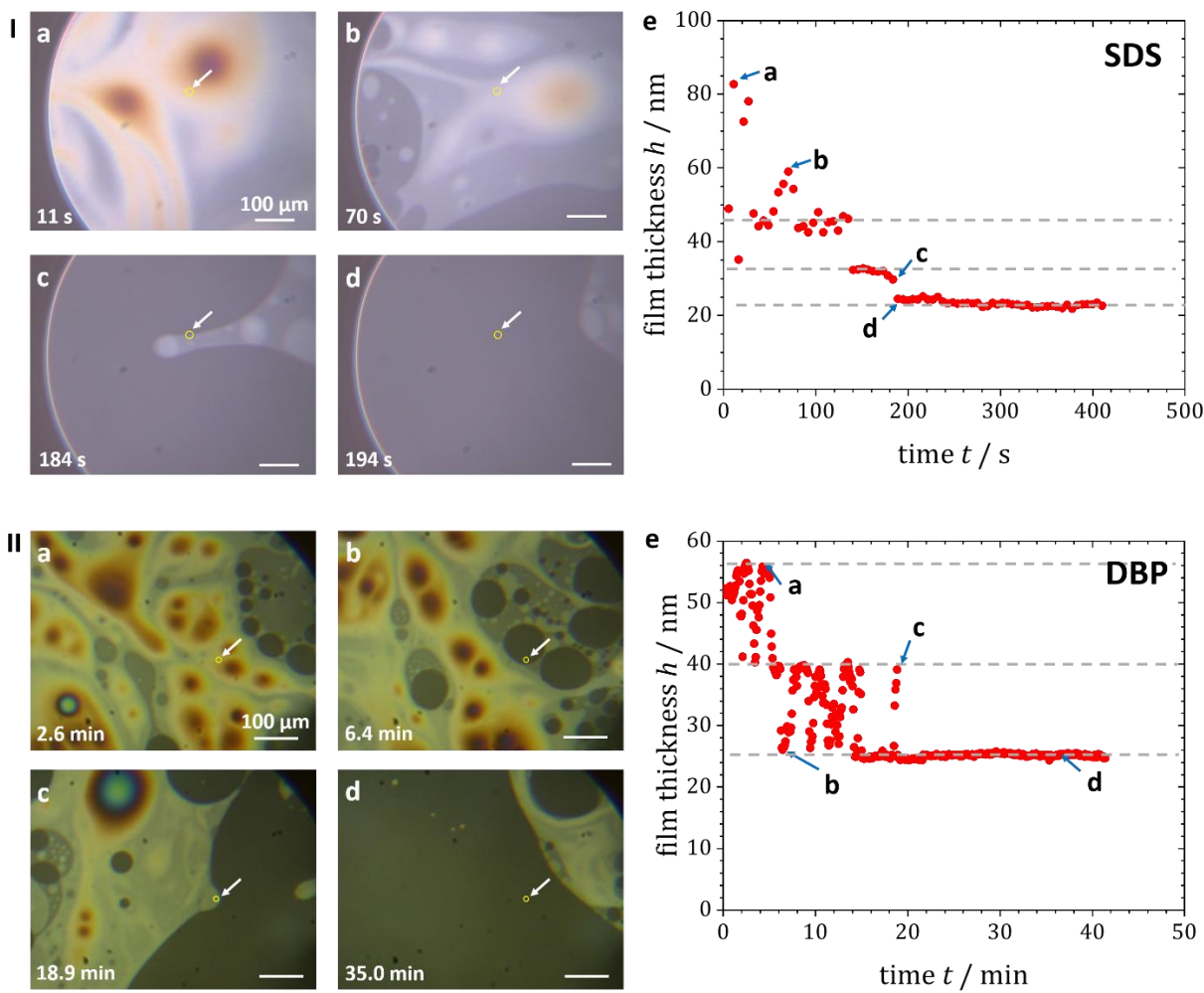

**Figure 4:** Photographs and film thickness analysis of films thinning over time observed using the µTFPB using two different solutions with very different viscosities ($10^{-3}$ vs 2.7 Pa s). I.a-d Photographs of an SDS film (at 0.24 mM) thinning under a capillary pressure of 100 Pa; and the corresponding film thickness over time (I.e) for the region displayed by the yellow rings and shown by the white arrows. II.a-d Pictures of a DBP-732 film thinning under a capillary pressure of 100 Pa; and the corresponding film thickness over time (II.e) for the region depicted by the yellow rings and shown by the white arrows.

### 3.2. High viscosity foam films (DBP-732 polymer melt)

Figure 4.II.a-d shows the drainage over time of the polymer melt DBP-732 subjected to a capillary pressure of 100 Pa monitored with the µTFPB. Figure 4.II.e shows the evolution of the film thickness with time in the region depicted by the yellow circle pointed by the white arrows in the corresponding images. The refractive index used to calculate the film thickness was $n_l$ = 1.446, which was given by the manufacturer (Gelest). Similar to films from an aqueous surfactant solution, one observes the temporary coexistence of regions of different, well-defined thicknesses (i.e., regions of different colours and grey levels), which progressively decrease with time. Here the strata do not correspond to micelles, but to the characteristic dimensions of the macromolecules, as explained in a previous work[47]. The thickness of each stratum is equal to 15 ± 2 nm. Note that the equilibrium thickness of 25 nm is reached after 20 min, which is a very slow drainage compared to aqueous soap films (the equilibrium thickness was reached after 3 min for the SDS solution). This slow drainage is attributed to the high viscosity (2.7 Pa s) of the DBP-732 melt. Such highly viscous liquids can be treated without any problem with the µTFPB.

### 3.3. Gelling foam films

We show in this section qualitative examples of the drainage of polymer films (with an initial viscosity of 0.42 Pa s) stabilised by Glucopon 600 CSUP (an alkyl polyglycoside from Cognis, now BASF) undergoing physical cross-linking. The alginate solution contains 1.0 wt % alginate (coined as "low viscosity alginate" by the supplier, Alfa Aesar). Calcium ions ($Ca^{2+}$) are used as cross-linkers and are brought into the solution via calcium chloride dihydrate (from Sigma Aldrich). However, to control cross-linking, we protect the calcium ions with a chelating agent, namely Ethylene glycol-O,O'-bis(2-aminoethyl)-N,N,N',N'-tetraacetic acid, 97% (EGTA). At a neutral pH, the EGTA binds the $Ca^{2+}$ ions, which are not available for cross-linking. Upon acidification, below a pH of ca. 4, the EGTA protonates itself and releases the calcium ions, which become available to cross-link the alginates[50]. The concentration of $Ca^{+2}$ ions and EGTA was the same for all the alginate films, namely 0.1 M. We use 1 wt % D-(+)-glucono-delta-lactone (GDL, from Sigma Aldrich) to acidify the solution and initiate cross-linking[51].

Figure 5.I.a-d shows the drainage over time of a non-gelling 1 wt % alginate/0.1 M $Ca^{2+}$/EGTA film stabilised with 0.5 wt % Glucopon 600 CSUP. The film does not gel, as no GDL was added to acidify the solution. One sees that at early times, i.e. after 4 min (Figure 5.I.a), the film shows concentric bright and colourful rings centred in the middle of the film, the colour sequence indicating that the film is thickest in its centre. This is a well-known phenomenon, coined as "dimple"[52]. After 7.2 min, one sees in Figure 5.I.b that darker spots of "common black films"[15,20] appear at the edge of the film, which correspond to much thinner areas. As the film keeps draining, the darker regions percolate and the bright and colourful area on the film declines, as observed in Figure 5.I.c at 8.5 min. An even darker region appears upon further drainage, as seen on the right of the film in Figure 5.I.d, at 11.3 min, corresponding to the formation of a "Newton black film" (hydrated surfactant bilayer) accompanied by the formation of small bright spots known as Rayleigh instabilities[53]. This instability is now reasonably well understood and results from the fact that the solution is expelled from this zone faster than it can be drained away by the remaining film. This leads to the formation of a bump around the film, which, beyond a critical height, breaks into smaller "drops", where the film is locally thick[24–26,53].

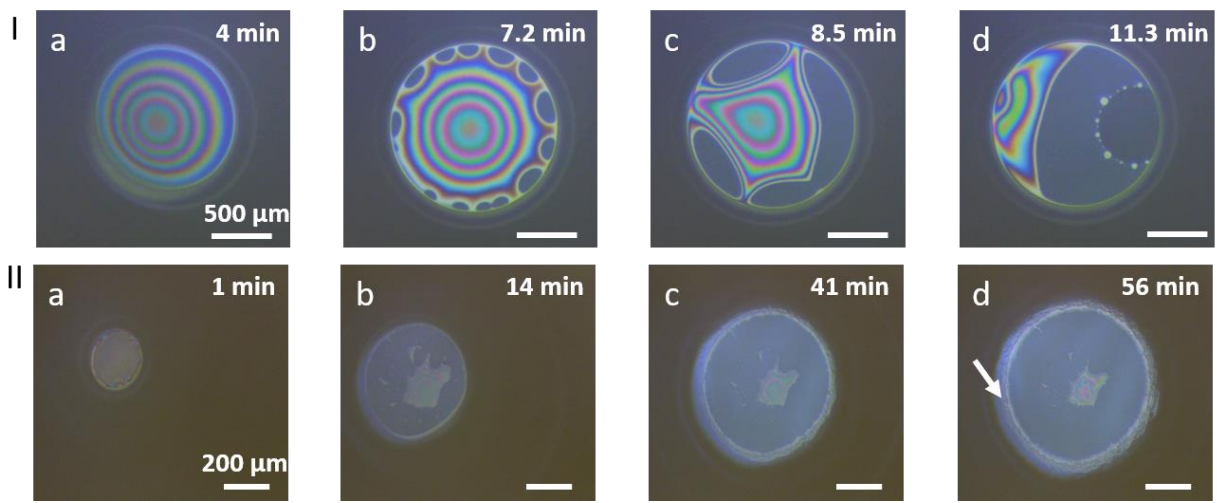

**Figure 5:** I.a-d) Photographs of a draining 1.0 wt % alginate film stabilised with 0.5 wt % Glucopon 600 CSUP without addition of an acidifying agent, i.e. the film does not undergo gelation and its bulk viscosity remains constant (0.42 Pa s) throughout drainage. II.a-d) Pictures of a draining 1 wt % alginate film stabilised with 0.5 wt % Glucopon 600 CSUP with addition of 1.0 wt % GDL to induce acidification which triggers cross-linking, i.e. gelation. The applied capillary pressure was 300 Pa.

Figure 5.II.a-d shows the drainage of a similar (yet gelling) 1 wt % alginate film with 0.1 M $Ca^{2+}$/EGTA stabilised with 0.5 wt % Glucopon 600 CSUP. Here we added 1 wt % of GDL to initiate cross-linking before film formation. Figure 5.II.a shows once again a drainage from the border of the film, leaving a dimple in its centre. However, after 14 min (Figure 5.II.b), we can observe a very different behaviour compared to the non-gelling film. The thickness of the overall film strops decreasing and the central area is not fully homogeneous but contains some zones which are significantly thicker than the surrounding film, as can be inferred from the colours. The precise origin of this phenomenon is not clear to us yet, but we suggest that it may result from a phase separation between polymer-rich globules and a polymer-poor film occurring as a result of confinement/drainage within the film. A more extensive and quantitative study is required to test this hypothesis. After 40 min (Figure *5*.II.c) and 56 min (Figure 5.II.d), the area of the flat film keeps increasing, whereas the area of the protruding globules keeps decreasing. The white arrow in Figure 5.III.e shows that the edge of the flat film is not only very well defined but also bounded by an even thinner zone. Such a film profile is hardly imaginable for a purely liquid film and must originate from the gelation of the film. Moreover, we could observe that the film was indeed gelled as we opened the chip to clean it. This is, to the best of our knowledge, the first example of the observation of drainage of a gelling film with the help of a thin film pressure balance. This experiment was made possible by the microfluidic chip which allows for a high entry pressures and an easy cleaning. Further qualitative studies are underway to understand the thinning behaviour observed for the gelling film, but a detailed discussion of the physical and physical-chemical processes at play during the drainage of gelling films is beyond the scope of the paper at hand.

## 4. Conclusions

Using a set of illustrative examples, we have introduced a new microfluidic Thin Film Pressure Balance (µTFPB) which allows to investigate the drainage and stability of free-standing liquid films in a reliable manner via the application of a well-defined capillary pressure. The main advantage of this design in comparison to previously used TFPBs is that it is suitable for a wide range of pressures and liquids, including highly viscoelastic or solidifying liquids. It therefore provides a convenient tool to investigate the evolution of solidifying films, for example, in order to understand the influence of interfaces/confinement on gelation and to explore the role of competing processes, such as drainage and gelation. The device can be fabricated using a micromilling machine and can be easily interfaced with the adequate flow/pressure control and optical techniques for the film analysis to fit the system studied and accuracy aimed for. The device design being very flexible, it can be adapted to a multitude of complementary investigations to study, for example, gas exchange or liquid/liquid films.

While the main focus of this article is on the demonstration of the device, the presented preliminary investigation of gelling free-standing films is the first of this kind. It is expected to open the pathway to a new field dedicated to the study of gelation under soft confinement and help shed light on the mechanisms responsible for pore opening in polymer foams.

## Authors' contributions

S. Andrieux wrote the first draft of the manuscript, carried out experiments, analysed the data, and participated in the development of the device. P. Muller and A. Cagna participated in the development of the device and proofread the manuscript. N.S. Macias Vera and M. Kaushal carried out experiments and proofread the manuscript. R. Bollache and C. Honorez helped design the device and mill the microfluidic chips. W. Drenckhan initiated and supervised the project.

## Conflict of interest

The authors have no conflict of interest to declare.

## Acknowledgment

We thank Cosima Stubenrauch, Patrick Kekicheff, Tetiana Orlova and Thibaut Gaillard for fruitful discussions. We also thank Aurélie Hourlier-Fargette for precious advice on alignment issues. We acknowledge funding from the Institute Carnot MICA (ThiFilAn project). We also acknowledge funding from the IdEx Unistra framework (Chair W. Drenckhan), from the European Research Council (ERC-METAFOAM 819511) and from the Institute of Advanced Studies at the University of Strasbourg (USIAS, Fellowship C. Stubenrauch).